\documentclass[
reprint, twocolumn, secnumarabic%
,amssymb,
amsmath,
nobibnotes, 
nofootinbib, aps, prl, showpacs,showkeys]{revtex4}

\usepackage{docs}%
\usepackage{bm}%
\usepackage{graphicx}
\usepackage{rotating}
\usepackage{hyperref}
\usepackage{url}
\usepackage{srcltx}
\expandafter \ifx \csname package@font\endcsname\relax\else
\expandafter \expandafter \expandafter
\usepackage
\expandafter \expandafter
\expandafter{\csname package@font\endcsname}%
\fi

\usepackage{epstopdf}



\usepackage{epstopdf}
\usepackage{enumerate}

\begin{document}

\title{The Proton Radius from Bayesian Inference}
\author{Krzysztof M. Graczyk}
\email{krzysztof.graczyk@ift.uni.wroc.pl}

\author{Cezary Juszczak}

\affiliation{Institute for Theoretical Physics, University of Wroc\l aw, pl. M. Borna 9,
50-204, Wroc\l aw, Poland}


\begin{abstract}
The methods of Bayesian statistics are used to extract the value of the proton radius
from the elastic $ep$ scattering data in a model independent way. 
To achieve that goal a large number of parametrizations 
(equivalent to neural network schemes) are considered and ranked by 
their conditional probability  $P(\mathrm{parametrization}\,|\,\mathrm{data})$ instead of using the minimal error criterion.
As a result the most probable proton radii values ($r_E^p=0.899\pm 0.003$ fm, $r_M^p=0.879\pm 0.007$ fm) are obtained and systematic error due to freedom in the choice of parametrization is estimated. 
Correcting the data for the two photon exchange effect leads to smaller difference  between the extracted values of $r_E^p$ and $r_M^p$. 
The results disagree with recent muonic atom measurements.
 
\end{abstract}

\maketitle
The problem of the proton radius is being discussed in the particle and atomic physics for many years but recently it has received even more attention due to the new results coming form the Lamb shift measurements of the muonic hydrogen atom ($\mu p$) \cite{Pohl:2010zza}. It is a very accurate  estimate of the proton radius but it is inconsistent with the CODATA value (compilation of the measurements in hydrogen atom and analysis of the electron-proton scattering data) \cite{Mohr:2012tt}. This disagreement, called ``the proton radius puzzle'',  has not been  explained yet.
There are many proposals to resolve the problem including
considering different types of interaction for $ep$ and $\mu p$ \cite{Kraus:2014qua}
or ``weakening the assumptions of perturbative formulation 
of quantum electrodynamics within the proton'' \cite{Pachucki:2014zea}.
In this paper we introduce a novel approach to the extraction of the proton radius from the $ep$ scattering data, which allows one to control the model dependence of the result within a Bayesian objective algorithm.  

The electromagnetic (E-M) structure of the proton is encoded in the electric $G_E$ and magnetic $G_M$ form factors, which are also the crucial input for the atomic (hydrogen-like) calculations \cite{Pachucki:1996zza}. In the low $Q^2$ limit in the Breit frame they are related with the distributions of the electric charge and magnetic momentum inside the nucleon \cite{Ernst:1960zza}. In particular the nucleon radius is expressed by the slope of the form factors at vanishing $Q^2$:
 \begin{equation}
\label{rp}
 r_{E,M}^p  \equiv \left( - \frac{6}{G_{E,M}(0)}\left.\frac{d G_{E,M}(Q^2)}{d Q^2} \right|_{Q^2=0} \right)^\frac{1}{2}.
\end{equation}

The value of $r_{E,M}^p$ can be obtained from  the  scattering data (mainly elastic $ep$).  Recent results include \cite{Bernauer:2010wm,Bernauer:2013tpr,Hill:2010yb,Lorenz:2012tm,Lorenz:2014vha} and a more complete list can be found in Refs. \cite{Epstein:2014zua,Karshenboim:2014vea,Kraus:2014qua}. 
These analyses have been performed in the spirit of the frequentistic statistics. 

The values of the proton radius (related to the electric charge distribution) obtained by different authors during the last fifty years,
range from about 0.8 fm to about 0.9 fm \cite{Sick:2003gm}. Indeed, as it was demonstrated in \cite{Hill:2010yb,Bernauer:2013tpr} the results depend the choice of the form factor parametrization but it is obvious that the collection of datasets used in the analysis also matters. 
Even the right choice of the number of parameters in a given class of parametrizations  can be a challenge. More parameters in the fit can always reduce the $\chi^2$ error but 
at some point the fit tends to reproduce statical fluctuations of the experimental measurements. Such a model overfits the data i.e.\ describes existing points with unrealistic precision but has no predictive power -- introducing new data points drastically increases the $\chi^2$ error and spoils the quality of the fit. 
This is connected with the so called bias variance trade-off.
Some attempt to resolve this problem has been made in Ref. \cite{Hill:2010yb} where conformal mapping was used
in order to exploit the analytic properties of the form factors.
\begin{figure}
\centering{
 \includegraphics[width=0.25\textwidth]{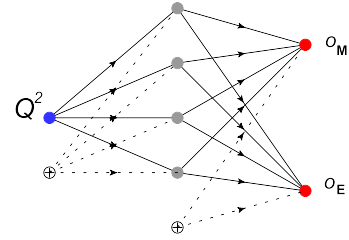}
 \caption{(Color online) The scheme of the neural network (with $H=4$ hidden units) used in the analysis, for reference see Eq.\ \ref{FF}. Every line (connection) corresponds to one parameter $w_i$. 
 Gray circles denote the sigmoid functions, open circles are constants.
 \label{Fig_network}}}
\end{figure}

Our philosophy is different. In contrast to the frequentistic statistics methods used in the above papers, we use the Bayesian statistics which is well suited to model comparison. 
Within this framework, adapted for neural networks, we select the most optimal model (according to the data) and extract the value of the proton radius. 

The extraction of the nucleon form factors in \cite{Graczyk:2010gw} was our first use of the Bayesian neural network framework. In the next paper \cite{Graczyk:2011kh} the approach was significantly improved to obtain the two-photon exchange (TPE) correction to unpolarized elastic $ep$ cross section. However, because of some physical assumptions these analyses were not dedicated to the low-$Q^2$ region. In the present paper we overcome this difficulty because:
 (i) the TPE contribution (elastic  and inelastic described by box diagrams with nucleon and $\Delta(1232)$ resonance as intermediate hadronic states) calculated in \cite{Graczyk:2013pca}  is subtracted from the data  and
 only E-M form factors are extracted from the cross section and polarization transfer data;
 (ii) the form factors are re-parameterized in order to automatically fulfil $G_E(0)=G_M(0)/\mu_p=1$ ($\mu_p$ is the proton magnetic moment);
 (iii) our software has been significantly improved in terms of accuracy and efficiency. 
  
Here we briefly outline the Bayesian framework,  for more details see \cite{Graczyk:2010gw,Graczyk:2011kh,Graczyk:2013pca}.
The main idea is to consider a large class of models. By a model we mean the function $\mathcal{N}$ used to fit the data $\mathcal{D}$ and two conditional probabilities: $P(\{w_i\}|\mathcal{N})$ -- prior which accommodates the initial assumption about the $\{w_i\}$, and the likelihood  $P(\mathcal{D}|\{w_i\}, \mathcal{N})$; $\{w_i\}$ is the set of parameters of the function $\mathcal{N}$. 
Then the models are ranked by
the conditional probability $P(\mathcal{N}|\mathcal{D})$, which 
 can be replaced by the evidence $P(\mathcal{D}|\mathcal{N})$
if the data is the same and no model is preferred at the beginning of the analysis.
\begin{figure}
\centering{
 \includegraphics[width=0.5\textwidth]{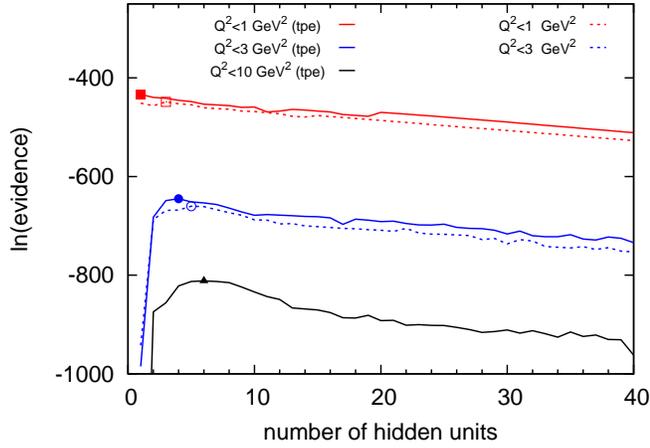}
   \caption{\label{Fig_evidence} (Color online) Logarithm of evidence for the best neural networks in each scheme. Solid/dashed lines correspond to the analyses of the data corrected/not corrected by the TPE effect.  Points mark the best models for each analysis.}}
\end{figure}

 For any given model we denote with $\{w_i\}_{MP}$ the configuration 
of the parameters which maximizes the posterior   
\begin{equation}
\label{posterior_paraeters_general}
P(\{w_i\}| \mathcal{D},\mathcal{N}) = \frac{P(\mathcal{D}|\{w_i \}, \mathcal{N})P(\{w_i \}|\mathcal{N})}{P(\mathcal{D}| \mathcal{N})}.
\end{equation}  
Finding $\{w_i\}_{MP}$ in one of the steps of the analysis.

Usually the evidence   is peaked at $\{w_i\}_{MP}$  and it simplifies to  the likelihood at the maximum  multiplied by the Occam factor which  makes too complex models  less likely \cite{MacKay_thesis}
\begin{equation}
P(\mathcal{D}|\mathcal{N}) \approx P(\mathcal{D}|\{w_i\}_{MP},\mathcal{N}) \underbrace{(2\pi)^\frac{W}{2} {(\det A)}^{-\frac{1}{2}}}_{Occam\,factor},
\end{equation}
where $W$ is the number of parameters $\{w_i\}$ and $A_{ij}= - \left. \nabla_{w_i}\nabla_{w_j} \ln P(\{w_i \}| \mathcal{D},\mathcal{N})\right|_{\{w_i \}=\{w_i \}_{MP}}$. 

For the function $\mathcal{N}$ we take the feed-forward neural networks with one hidden layer of units (Fig. \ref{Fig_network}), which correspond to linear combinations of sigmoid functions. This class of functions can be used to approximate any continuous function with arbitrary precision \cite{Cybenko_Theorem}.

The prior function for the neural network parameters $\{w_i\}$ has the standard form
$ \mathcal{P}\left(\{w_i\}\right|\left. \alpha, \mathcal{N}\right)
\sim  {e^{- \alpha E_w}}$  $\mathrm{where}\,E_w=\frac{1}{2}\sum_{i=1}^{W}w_i^2$.
The parameter $\alpha$ is the regularizer which determines the width of the initial Gaussian distribution for the parameters. In principle it should be treated as another parameter of the model and its optimal value $\alpha_{MP}$ 
is also obtained during the analysis.
\begin{figure}
\centering{
 \includegraphics[width=0.5\textwidth,height=6.52cm]{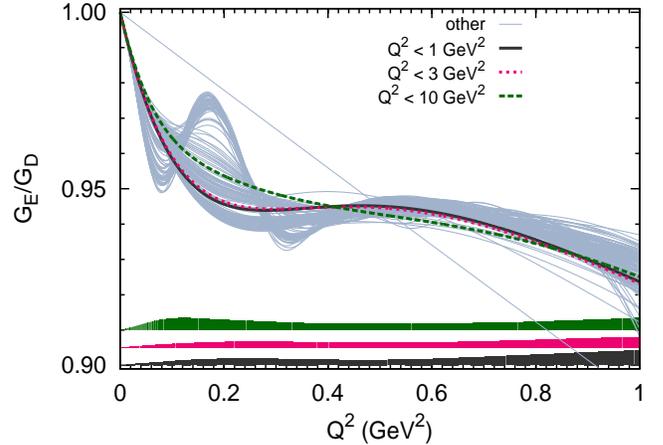}
  \caption{(Color online) The form factor $G_E/G_D$ ($G_D = 1/(1 +Q^2/0.71 
  \mathrm{GeV}^2)^2$). Red, blue and magenta lines correspond to the best models
  in the analysis with data limited by $Q^2_{\mathrm{cutoff}}=1$, $3$ and $10$ GeV$^2$ respectively. 
The areas colored with magenta, blue and red denote $1\sigma$ uncertainty due to the change in the fit parameters (calculated within the Hessian approximation). \label{Fig_GE}   The gray lines show models which are best for neural networks with definite number of hidden units, but are not globally best.
}}
\end{figure}
The likelihood is defined by the $\chi^2$ distribution
\begin{equation}
\mathcal{P}\left(\mathcal{D}\right|\left.\{w_i\}, \alpha, \mathcal{N} \right) \sim  e^{-\chi^2_{cr.}(\mathcal{D},\{w_i\}) - \chi^2_{PT}(\mathcal{D},\{w_i\})}.
\end{equation} 
The $\chi_{cr.}^2$ includes the unpolarized elastic $ep$ scattering cross section  data, which are assumed to be parametrized by one-photon exchange formula
$
\sigma_R= Q^2 G_M^2 /4M_p^2 + \varepsilon G_E^2
$ ($M_p$ is the proton mass, $\varepsilon$ denotes photon polarizability).
Each independent set of cross section measurements is characterized by its systematic normalization uncertainty so we introduce a separate normalization parameter for every data set. Values of these parameters are obtained by an iterative algorithm described in   \cite{Graczyk:2011kh}. The $\chi^2_{PT}$ includes the form factor ratio $\mu_p G_E/G_M$ data from the polarization transfer measurements. We consider the same selection of data as in \cite{Graczyk:2011kh} with one additional data set from \cite{Zhan:2011ji} and make the prior assumption that all data sets are equally relevant.  
 The MAMI  data \cite{Bernauer:2010wm} are not included in the analysis because of the difficulties described in \cite{Arrington:2011kv}.

The optimal configuration $\{\alpha, w_{i}\}_{MP}$ is 
found by an iterative procedure based on the fact that
the maximum of the posterior corresponds to the minimum over $\{w_i\}$ of the error function 
$S(\mathcal{D},\{w_i\})\equiv \chi^2_{cr.}(\mathcal{D},\{w_i\})+\chi^2_{PT}(\mathcal{D},\{w_i\}) + \alpha_{MP} E_w$ and also
$\left.\frac{\partial }{\partial \alpha} \mathcal{P}\left(\mathcal{D}\right|\left. \alpha, \mathcal{N} \right)\right|_{\alpha=\alpha_{MP}} =0$,
where $$ \mathcal{P}\left(\mathcal{D}\right|\left. \alpha, \mathcal{N} \right) = \int d w \mathcal{P}\left(\mathcal{D}\right|\left.\{w_i\}, \alpha, \mathcal{N} \right)
\mathcal{P}\left(\{w_i\}\right|\left. \alpha, \mathcal{N}\right). $$

Eventually  the logarithm of evidence for the model $\mathcal{N}$ (in the Hessian approximation) reads:
\begin{eqnarray}
\ln P(\mathcal{D}| \mathcal{N}) & \approx &-S(\mathcal{D},\{w_i\}_{MP}) + \frac{W}{2}\ln \alpha_{MP} - \frac{1}{2}\ln \det A \nonumber \\
& &   - \frac{1}{2}\ln\sum_{i=1}^W \frac{1}{2}\frac{\lambda_i}{\lambda_i +\alpha_{MP}} 
\end{eqnarray}
where $A_{kj} = \nabla_{w_k}\nabla_{w_j} S(\mathcal{D},\{w_i\}) $ and $\lambda_k$-s are the eigenvalues of the matrix $\nabla_{w_i}\nabla_{w_j}(\chi_{cross}(\mathcal{D},\{w_i\}) + \chi_{PT}(\mathcal{D},\{w_i\}))$.

As the form factors $G_E$, $G_M$ describe the properties of the same object, they must be related, so we parametrize them with a single feed forward neural network with one input ($Q^2$) and  two outputs, $o_E$ and $o_M$ (see Fig. \ref{Fig_network}), 
\begin{equation}
\label{FF}
G_{E} = (1  - Q^2 o_E) G_D, \quad G_{M} = \mu_p (1  - Q^2 o_M) G_D,
\end{equation}
where $G_D(Q^2) = \left(1 + {Q^2}/{M_V^2} \right)^{-2}$ and $M_V^2=0.71$ GeV$^2$.
Then the value of the proton radius is given by
\begin{equation}
\label{r2}
\left(r^p_{E,M}\right)^2 =  \left(r^p_{dipole}\right)^2  + 6 o_{E,M}(Q^2=0),
\end{equation}
where ${(r_{dipole}^p})^2 = 12/M_V^2$. We see that the network outputs at $Q^2=0$ directly contribute to the deviation of the proton radius from the dipole form factor value.

In Refs. \cite{Rosenfelder:1999cd,Sick:2003gm} it was shown that modifying the cross section data by the Coulomb distortion (CD) is important in the extraction of the proton radius. On the other hand, the CD is a part of the TPE effect\footnote{The relevance of the TPE effect and  higher order Born corrections in  the proton radius extraction is also discussed in Refs. \cite{Gorchtein:2014hla} and \cite{Arrington:2012dq,Arrington:2004is} respectively.}, which has to be subtracted from the data to obtain the form factors which agree with the polarization transfer measurements \cite{Arrington:2007ux}. Similarly as in  \cite{Blunden:2005jv}  we correct the cross section data by the TPE contribution (calculated in  \cite{Graczyk:2013pca} in similar way as in \cite{Blunden:2005ew,Kondratyuk:2005kk}).

The set of all possible parametrizations is infinite. In our approach the models are grouped according to the number of units in the hidden layer. For each group we found the  $\{w_i,\alpha\}$ which maximize the evidence. The dependence of the maximal  evidence on the number of hidden units is plotted in Fig. \ref{Fig_evidence}. For every  analysis the evidence reaches a peak (which defines the best model) and then starts to fall. Because of this fact we do not consider the networks with more then 40 hidden units. The total number of obtained neural network parameterizations exceeded half million. 

It needs to be underlined that the models which maximize the evidence are not those which minimize the $\chi^2$. This is true in each class of models and also globally. The $\chi^2_{min}$ decreases with the number of parameters possibly saturating at some point. But the models with the lowest $\chi^2$ tend to  overfit the data, which is one of the reasons why the $\chi^2_{min}$ is not a suitable criterion for ranking the models. The other problems of $\chi^2$-based criteria are discussed in \cite{Kraus:2014qua}. Objective mathematical methods for model comparison
are provided by the Bayesian statistics.

\begin{figure}
\centering{
 \includegraphics[width=0.55\textwidth]{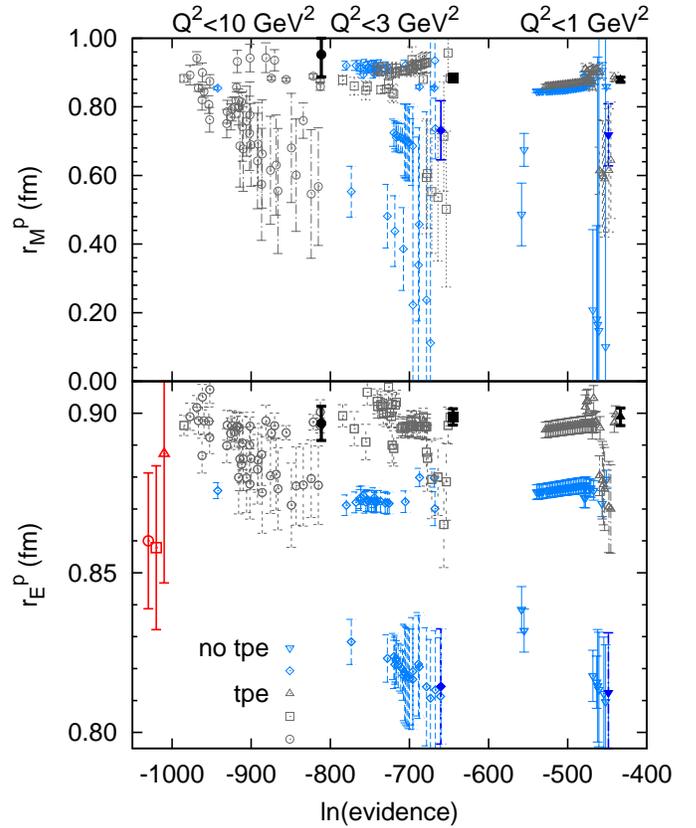}
   \caption{(Color online) The proton radii corresponding to the models which are the best within particular data selection and neural network scheme. The results for the data corrected by the TPE 
   are marked with black $\triangle$,  $\square$, and $\bigcirc$ for $Q^2_{\mathrm{cutoff}}=1$, $3$ and $10$   GeV$^2$ respectively. 
   The results for the uncorrected data  are marked with blue $\nabla$ and $\Diamond$  for $Q^2_{\mathrm{cutoff}}=1$ and $3$ GeV$^2$ respectively.
The filled points mark the best result for each case. 
The three leftmost red points are the best results according to the minimum of the error function. 
For clarity their x-coordinate has been changed. Their ln(evidence) values are $-718$, $-1070$, and $-1311$.
The picture is based on the analysis of $39\cdot 5 =195$ models.\label{Fig_proton_radius}}}
\end{figure}

In Fig. \ref{Fig_GE} it is shown  how the form factor plots depend on the choice of the parametrization. The gray lines correspond to the best models in each group, while the globally best for each analysis are shown with color.
Certainly the choice of the form factor parametrization has also strong impact on the obtained proton radius value (Fig. \ref{Fig_proton_radius}). Hence it is crucial to have criterion for finding the model which is the most favorable by the measurements.  On the other hand each model can be true with the probability $P(\mathcal{N}|\mathcal{D})$ (proportional to the evidence) so it is possible to calculate the expected value of the proton radius and the systematic uncertainty due to the choice of the model. In practice the expected value is very close to the most probable value (see Tables \ref{Tab_radii} and \ref{Tab_radii_model}).
\begin{table}

\begin{tabular}{|c|c|c|c|}
\hline
\hline
$Q^2_{cutoff}$ (GeV$^2$) & $r_{M}^p$ (fm) & $r_{E}^p$  (fm) & $H$\\
\hline
$ 1$       & $ 0.879	\pm 0.007 $  & $0.899 \pm	0.003$ & 1\\
$3$       & $0.883 \pm 	0.007$	& $0.899	\pm 0.003$ & 4\\
$10$      & $0.953	\pm 0.065$ & $0.897 \pm	0.005$ & 6 \\
\hline
\hline
\end{tabular}
\caption{The values of the proton radii  obtained with $1\sigma$ uncertainty due to variation in the parameter space. $H$ is the number of hidden units of the best model.\label{Tab_radii}}
\end{table}

\begin{table}
\begin{tabular}{|c|c|c|}
\hline
\hline
$Q^2_{cutoff}$ (GeV$^2$) & $r_{M}^p$ (fm) & $r_{E}^p$  (fm) \\
\hline
$1$        & $ 0.8792 \pm	0.0006 $  & $0.8989	\pm 0.0001$ \\
$3$       & $0.8828 \pm	0.0063$	& $0.8988	\pm 0.0003$ \\
$10$      & $0.9205 \pm	0.0606$ & $0.8968   \pm	0.0029$ \\
\hline
\hline
\end{tabular}
\caption{The expected value (according to evidence probability distribution) of the proton radii with systematic uncertainty due to the choice of the parametrization (given by the variance).\label{Tab_radii_model}}
\end{table}

In Fig. \ref{Fig_proton_radius} we show the proton radii values obtained from the analyses where data points with $Q^2$ above $Q^2_{\mathrm{cutoff}}=1$, $3$ and $10$ GeV$^2$ are rejected (for the data uncorrected by the TPE  we considered only $Q^2_{\mathrm{cutoff}}=1, \; 3$ GeV$^2$). The choice of the value for the cutoff  has small impact on the extraction of the proton radii.  For the result of our analysis we take those from the lowest cut, namely
$r_{E}^p=0.899\pm 0.003$ fm and $r_{M}^p=0.879\pm 0.007$ fm. On the other hand the $r_E^p$'s obtained based on the minimal error criterion depend on the cutoff choice. These results are  smaller than those indicated by the evidence and have larger uncertainties, see Fig. \ref{Fig_proton_radius}. We leave for future studies within the Bayesian framework the quantitative investigation of the dependence of the results on the data set selection and $Q^2_{\mathrm{cutoff}}$ value. Eventually it can be seen in Fig.\ \ref{Fig_proton_radius}  that using the data uncorrected by TPE effect leads to larger uncertainties of  $r_E^p$ and $r_M^p$ and larger difference  $r_E^p-r_M^p$ ($0.02$ fm with TPE and $0.09$ fm without). 

We would like to emphasize that according to our knowledge the present work is the first extraction of the proton radius based on the Bayesian methods.
The extracted value of the proton radius $r^p_E$ agrees with previous results \cite{Sick:2003gm,Blunden:2005jv,Hill:2010yb} but disagrees with the muonic atom measurements  \cite{Pohl:2010zza}. The subtracting of the TPE correction from the cross section data plays an important role in the analysis.

\vspace*{1mm}

The calculations have been carried out in Wroclaw Centre for
Networking and Supercomputing (\url{http://www.wcss.wroc.pl}),
grant No. 268.

\bibliographystyle{apsrev4-1}
\bibliography{bibdrat}

\end{document}